\documentclass[conference]{IEEEtran} % use the `journal` option for ITherm conference style
\IEEEoverridecommandlockouts
\usepackage{cite}
\usepackage{amsmath,amssymb,amsfonts}
\usepackage{algorithmic}
\usepackage{graphicx}
\usepackage{subcaption}
\usepackage{textcomp}
\usepackage{xcolor}
\usepackage{multirow}
\usepackage[normalem]{ulem}
\usepackage{flushend}
\usepackage{tikz}

\useunder{\uline}{\ul}{}
\def\BibTeX{{\rm B\kern-.05em{\sc i\kern-.025em b}\kern-.08em
    T\kern-.1667em\lower.7ex\hbox{E}\kern-.125emX}}
\begin{document}

\title{Development and Validation of a Proximity-based Wearable Computing Testbed for Community-oriented Wearable Systems
}

\author{\IEEEauthorblockN{1\textsuperscript{st}Qimeng Li}
\IEEEauthorblockA{\textit{DIMES} \\
\textit{University of Calabria}\\
Rende, Italy \\
qimeng.li@unical.it}\\

\IEEEauthorblockN{4\textsuperscript{th}Raffaele Gravina}
\IEEEauthorblockA{\textit{DIMES} \\
\textit{University of Calabria}\\
Rende, Italy \\
r.gravina@dimes.unical.it}

\and
\IEEEauthorblockN{2\textsuperscript{nd}Fabrizio Mangione}
\IEEEauthorblockA{\textit{DIMES} \\
\textit{University of Calabria}\\
Rende, Italy \\
mngfrz00c10d086p@studenti.unical.it}\\

\IEEEauthorblockN{5\textsuperscript{th}Giancarlo Fortino}
\IEEEauthorblockA{\textit{DIMES} \\
\textit{University of Calabria}\\
Rende, Italy \\
giancarlo.fortino@unical.it}
\and
\IEEEauthorblockN{3\textsuperscript{rd}Francesco Porreca}
\IEEEauthorblockA{\textit{DIMES} \\
\textit{University of Calabria}\\
Rende, Italy \\
prrfnc98h18d086s@studenti.unical.it }\\
}

\maketitle
\begin{tikzpicture}[remember picture,overlay,shift={(current page.north)}]
    \node[anchor=north,yshift=0.1cm]{\includegraphics[width=17cm]{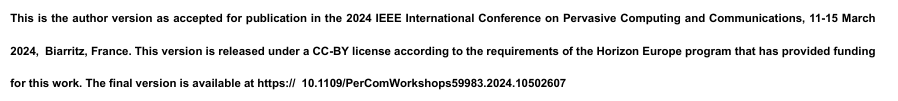}};
\end{tikzpicture}

\begin{abstract}
In the rapidly evolving digital technology landscape, community-oriented wearable computing systems are emerging as a key tool for enhancing connectivity and interaction within communal spaces. This paper contributes to this burgeoning field by presenting the development and implementation of a proximity-based wearable computing testbed designed to forge stronger links within communities. The testbed exploits Ultra-Wideband (UWB) position sensors, 9-axis motion sensors, edge nodes, and a centralized server, forming a cohesive network that actively facilitates community interactions and engagements. By employing anchors and targets within the UWB sensors, the system achieves high precision in location and distance measurements, laying the groundwork for various proximity-based applications. Integrating 9-axis motion sensors and advanced edge nodes further underscores the system's versatility and robustness in wearable and edge computing. This paper delves into an in-depth exploration and evaluation of the proposed system's architecture, design, and implementation processes. It provides a comprehensive analysis of experimental results and discusses the system's potential impact on enhancing community networks, along with the future directions this technology could take.
\end{abstract}


\section{Introduction}
\label{Section 1}

The advent of wearable computing~\cite{d2022situation} signifies a pivotal shift in human-technology interaction, revolutionizing how individuals connect within their communities. These technologies facilitate personal connectivity and play a crucial role in strengthening communal bonds and enriching collective experiences.

Our previous research~\cite{li2019group} highlighted the limitations of using motion sensors in wearable devices, which provided insights into individual movement patterns but fell short in identifying group dynamics or shared activities. To address this, we introduce the concept of proximity or spatial closeness in our current study. Human interactions often occur within a specific range of proximity, making it a critical factor in understanding and enhancing communal activities.

This paper introduces an innovative proximity-based wearable computing testbed. It is designed to transform community interactions and engagements by leveraging the synergy of advanced technologies. At its core, the system integrates Ultra-Wideband (UWB) position sensors~\cite{huang2021ultra} for precise location tracking and 9-axis motion sensors~\cite{li2019group, ma2019embedded} to capture intricate movement details. Supported by edge computing, this network excels in accuracy and responsiveness, which is essential for community-centric applications.

We demonstrate the testbed's practicality in everyday contexts, showcasing its seamless integration into communal life and its potential to make these spaces more interactive, responsive, and connected. The paper comprehensively details the testbed's architecture, design, implementation, and evaluation, emphasizing its technical sophistication and contribution to community networks. Through rigorous real-world evaluations, we illustrate the testbed's significant impact on enhancing community interactions, thereby underscoring its potential to revolutionize communal technological landscapes.

Furthermore, it is important to note that our work is still nascent. We have successfully established the foundational framework of the platform and conducted preliminary simulations to test its functionality. These initial experiments serve as a crucial stepping stone in understanding the dynamics of wearable computing in community settings. As such, while our results are promising, they represent the early phases of a much broader research trajectory aimed at fully realizing the potential of this technology in communal environments.
%The remainder of this paper is organized as follows: Section 2 reviews related works, positioning our contributions within the existing landscape of technological advancements. Section 3 elucidates the system architecture, detailing the components and their interplay. Subsequent sections navigate through the design, implementation, experimental evaluation, and discussion, culminating in a conclusion that encapsulates the essence of our research findings and future prospects.

\section{Related Work}
Various studies have explored the domain of wearable computing and proximity-based systems, particularly focusing on community-oriented applications and interactions. This section presents an overview of the relevant literature, categorizing the discussed works into proximity-based systems in communities and innovations in wearable computing.

\subsection{Proximity-based Systems in Communities}

Proximity-based systems have significantly transformed community networks, bolstering interaction and connectivity. Yu et al.~\cite{6569419} introduced a novel method for detecting and tracking context-aware communities within proximity-based mobile social networks. By employing an influence graph, their approach integrates social contexts into user contacts, thereby enriching the mobile social networking experience and enhancing application performance.

Advaith et al.~\cite{9198394} proposed a unique technique for community detection in dynamic social networks. Their method converts temporal graphs into static ones while retaining crucial temporal data. This innovation allows using conventional community detection algorithms, significantly reducing computational time for large-scale, complex graph analyses.

The authors discuss a practical implementation of proximity-based systems in public health in~\cite{9315498}, particularly focusing on COVID-19 contact tracing in densely populated workplaces. This system effectively identifies infection risks using Bluetooth Low Energy (BLE) technology for proximity data collection, thereby providing an efficient alternative to conventional mass testing approaches. This digital tracing solution is particularly advantageous in heavily populated areas with strained healthcare resources, illustrating the critical role of proximity-based systems in modern public health management.

\subsection{Wearable Computing Innovations}
Wearable computing has spurred transformative advancements, particularly in health monitoring, social interaction, and public safety. A notable development by Bian et al.~\cite{10.1145/3410531.3414313} is a wearable proximity sensor for enhancing social distancing during the COVID-19 pandemic. Using oscillating magnetic fields, this system outperforms traditional Bluetooth-based technologies in accuracy and robustness. Another significant innovation is 'FedHealth~\cite{9076082}', a federated transfer learning framework designed for personalized healthcare in wearable devices. It addresses data isolation and personalization challenges effectively, demonstrating its utility in applications like activity recognition and Parkinson's disease diagnosis without compromising user privacy. These advancements highlight the growing role of wearable computing in addressing critical health issues and enhancing user experiences in various domains.

%\subsection{UWB Position and Edge Computing}

%Ultra-wideband (UWB) position sensors have been pivotal in enhancing positioning and location-based services.

%Concurrently, edge computing has emerged as a crucial component in managing and processing data in proximity-based systems.

\section{System Architecture}

The architecture of our proximity-based wearable computing testbed is designed with a focus on fostering seamless community connectivity. It is comprised of Ultra-Wideband (UWB) position sensors, 9-axis motion sensors, Raspberry Pi edge nodes, a centralized server, and other essential components such as Keepalived, HAProxy ~\cite{mihai2023integrated}, and MQTT Broker. This section delves into the architectural components and their interrelationships within the system.

\subsection{Testbed Specifications}

The specifications of the testbed are designed to ensure precision, reliability, and efficiency in data collection and processing. The testbed is characterized by its functional diversity and the specific technical capabilities of its components.

\paragraph{Data Computation and Network Management}
The system's core, responsible for data computations and network connectivity, is powered by devices like the Raspberry Pi 4 Model B. This node has a Broadcom BCM2711, Quad-core Cortex-A72 (ARM v8) 64-bit SoC, and up to 8GB RAM, ensuring robust computing power. It offers high-speed wireless and wired networking capabilities (including WiFi, BLE, and LAN), facilitating seamless data transmission within the testbed network.

\paragraph{Positioning and Tracking}
For precissystem's backboneking, the ESP32 UWB Pro is utilized. Specialized in ultra-wideband positioning, this node is instrumental for accurate distance measurement and location tracking. It employs Time of Flight (ToF) localization methods to deliver high precision and reliability, crucial for proximity-based applications, with the capability to determine distances within a few centimeters (10-30cm).

\paragraph{Motion Sensing and Data Integration}
The BNO055 9-DOF inertial measurement unit (IMU), a 9-axis motion sensor encompassing a gyroscope, accelerometer, and magnetometer, is employed for comprehensive motion tracking. It provides integrated data like orientation and linear acceleration, essential for dynamic interaction analysis in wearable computing. This sensor is especially suitable for applications requiring detailed user movement and orientation data, such as gesture recognition and activity monitoring.

These nodes form the backbone of the system, each contributing to the testbed's capabilities. Their integration and coordinated functions enable the testbed to deliver high precision, efficiency, and flexibility in wearable computing applications.

\subsection{Node Arrangement and Connectivity}
\begin{figure} [!ht]
    \centering
    \includegraphics[width=1\columnwidth]{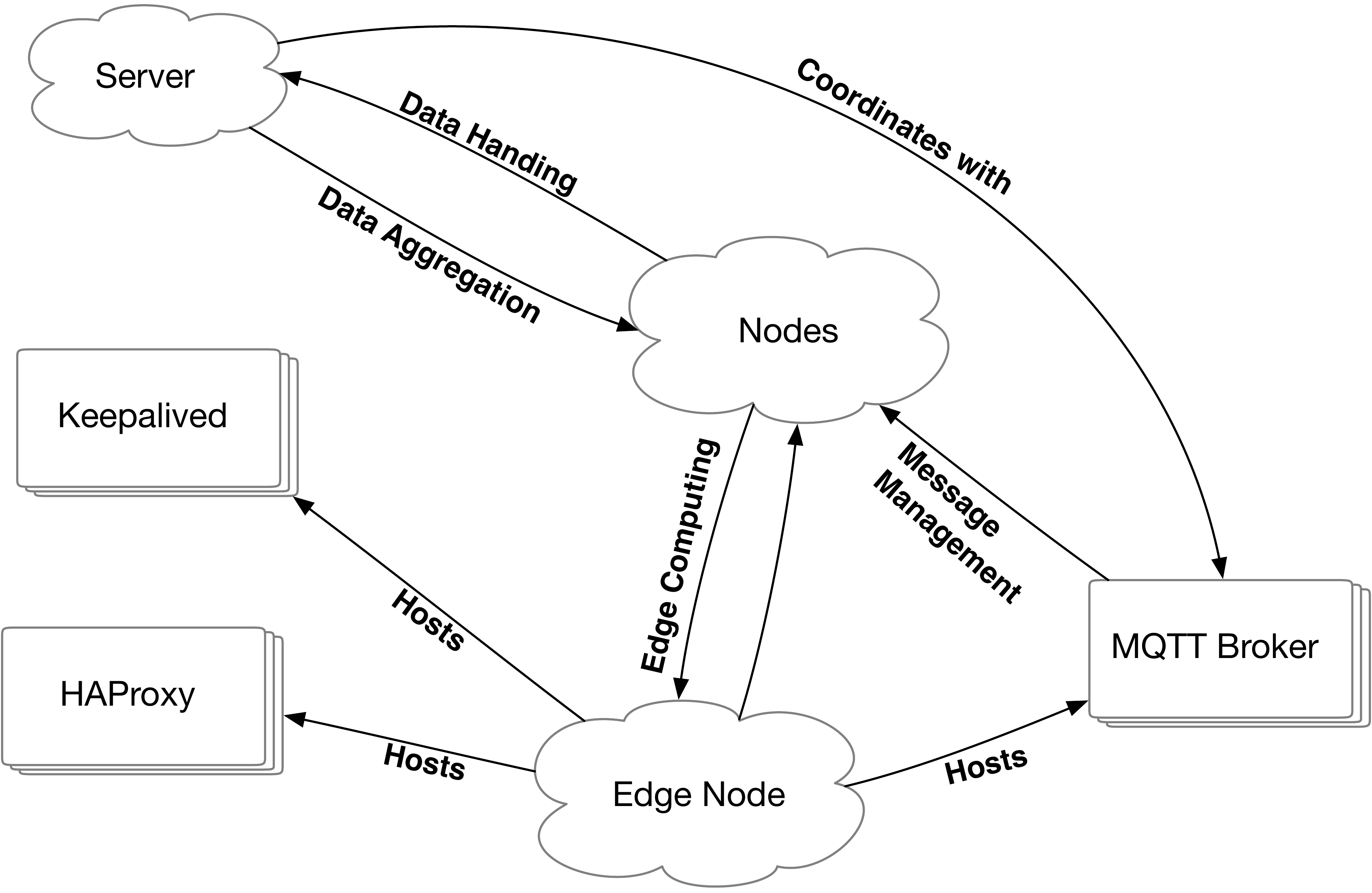}
    \caption{Node Arrangement and Connectivity.}
    \label{fig:conectivity}
\end{figure}

The arrangement and connectivity of nodes in the testbed (see Figure~\ref{fig:conectivity}), along with the integration of Keepalived, HAProxy, MQTT Broker, and the use of Anchor and Tag nodes, form a comprehensive network system that ensures reliability, efficiency, and scalability.

\begin{itemize}
    \item \textbf{Keepalived and HAProxy:}
    \begin{itemize}
        \item \textit{Purpose:} Ensure high availability and failover management (Keepalived) and efficient load balancing (HAProxy) across the nodes.
        \item \textit{Integration:} These tools are integrated to maintain uninterrupted service and optimal load distribution, critical for real-time data processing and system uptime.
    \end{itemize}

    \item \textbf{MQTT Broker:}
    \begin{itemize}
        \item \textit{Purpose:} Serves as the message broker in the MQTT protocol, orchestrating communication between different nodes.
        \item \textit{Integration:} The broker is integrated to ensure efficient, reliable data transmission and correct message publication and subscription by relevant nodes.
    \end{itemize}

    \item \textbf{Tag Nodes:}
    \begin{itemize}
        \item \textit{Purpose:} Primarily responsible for communicating position data within the UWB system. They enable dynamic location tracking of objects or individuals, crucial for real-time monitoring and interaction analysis.
        \item \textit{Integration:} Tags are directly connected to sensing nodes that collect IMU data. This integration allows for the seamless transmission of both IMU and positional data to the data processing center. This setup not only facilitates real-time tracking but also enriches the data with detailed motion and orientation information, significantly enhancing the depth and utility of the collected data within the testbed environment.
    \end{itemize}

    \item \textbf{Anchor Nodes:}
    \begin{itemize}
        \item \textit{Purpose:} Serve as fixed reference points within the UWB system to provide accurate positioning and establish the spatial framework of the testbed.
        \item \textit{Integration:} Anchors are strategically placed and integrated to create a stable grid for precise location tracking, enhancing the system's accuracy in proximity-based applications.
    \end{itemize}

    \item \textbf{Server:}
    \begin{itemize}
        \item \textit{Purpose:} Manages data aggregation and processing, while coordinating with Keepalived and HAProxy for system reliability.
        \item \textit{Integration:} The server is integrated to work in tandem with the MQTT Broker, facilitating efficient data flow and network communication.
    \end{itemize}
\end{itemize}

This sophisticated arrangement and interconnectivity of the various components ensure that the testbed is efficient in data handling and processing and robust regarding system availability, reliability, and scalability.

\subsection{Programming Environment and Capabilities}
\begin{itemize}
    \item \textbf{Arduino IDE:} Utilized for programming the ESP32 UWB Pro nodes, this IDE enables the implementation of complex functionalities such as distance sensing and precision tracking. It provides a user-friendly platform for developing and uploading code to ESP32 modules.

    \item \textbf{PyCharm IDE:} PyCharm IDE is used for developing Python scripts and applications for the Raspberry Pi nodes. It supports the creation of custom algorithms essential for processing data collected by the system. PyCharm enhances the development process with features like code analysis, graphical debugging, and version control integration, facilitating efficient real-time analysis and decision-making.
\end{itemize}

This programming environment is integral to the testbed, providing the necessary tools for developing and deploying the software that drives its various functionalities. The combination of Arduino IDE and PyCharm IDE ensures a robust and versatile development ecosystem, catering to the diverse programming needs of the testbed's components.

\subsection{System Software and Data Processing}
The testbed's system software includes a centralized server, an MQTT broker, and an edge computing layer on Raspberry Pi units, each playing a critical role in the testbed's operation.

The centralized server is at the heart of the testbed's data management strategy. It hosts a robust data storage system and it is responsible for advanced data processing tasks like analytics, pattern recognition, and machine learning. This server is designed to be scalable, allowing for increased computational resources and storage capacity as needed.

An MQTT broker, a central component of the MQTT protocol, manages communication between the nodes and the server. This broker ensures reliable and efficient message delivery, which is essential for maintaining real-time data communication integrity. Its lightweight and scalable nature makes it particularly well-suited for IoT environments like this testbed.

Edge computing is implemented through Raspberry Pi units strategically placed within the testbed. These units process data locally, which reduces latency and bandwidth requirements for data transmission to the server. They enable real-time analytics and decision-making at the network's edge, enhancing system responsiveness. The distributed nature of this computing architecture also improves system resilience and supports parallel data handling and analysis.

%\subsection{Application Overview}
%The testbed supports applications such as proximity-based activity recognition and predictive control. These applications demonstrate the system's utility in creating interactive and connected community environments.

%\subsection{Experimentation and Data Generation Process}
%The testbed is designed for rigorous experimentation. It includes:
%\begin{itemize}
%    \item Procedures for setting up and conducting experiments.
%    \item Data generation for machine learning models and simulation parameterization.
%\end{itemize}

%This system architecture represents a significant advancement in wearable computing, enabling new forms of connectivity and interaction in community networks.

\section{Experiments and Evaluation}

\subsection{Experimental Setup}
\begin{figure} [!ht]
    \centering
    \includegraphics[width=1\columnwidth]{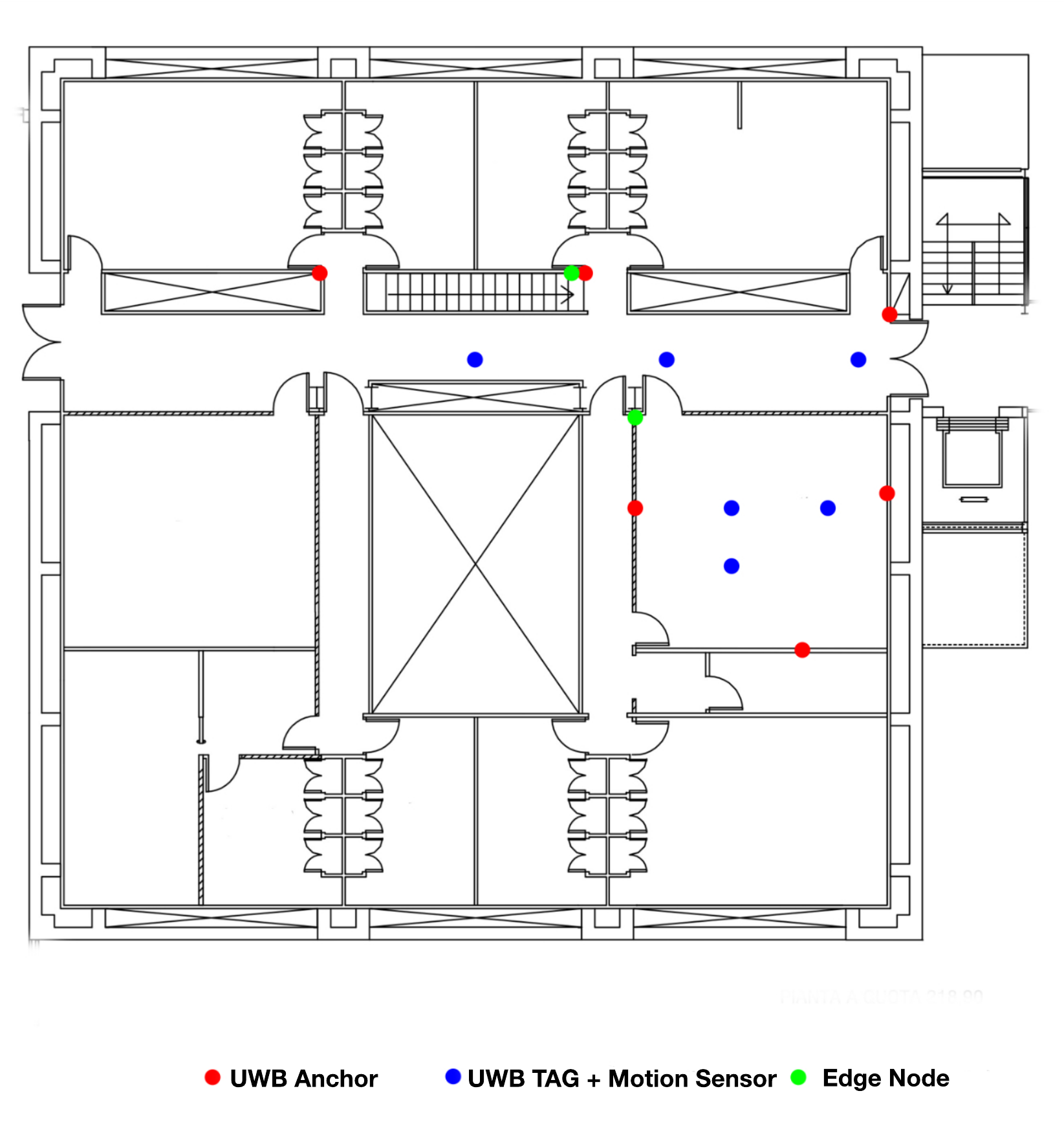}
    \caption{Experiments Environment and Sensor deployment.}
    \label{fig:exp_enviro}
\end{figure}

The experimental setup is meticulously designed to emulate real-world scenarios, enabling an in-depth performance evaluation of the wearable computing system. The setup incorporates Raspberry Pi units, ESP32 UWB Pro nodes, and BNO055 sensors to ensure comprehensive data accuracy and optimal coverage. Data collected include positional and motion metrics, forming a rich dataset for subsequent analysis (see Figure~\ref{fig:exp_enviro}).

\begin{itemize}
    \item \textbf{Test Environment:} The experiments are conducted in a controlled environment that simulates real-world conditions, allowing for collecting relevant data and observing system interactions under varied circumstances.

    \item \textbf{Node Deployment:} Raspberry Pi units, ESP32 UWB Pro nodes, and BNO055 sensors are strategically placed to ensure optimal coverage and data accuracy. The placement of Anchor and Tag nodes is particularly crucial for precise UWB positioning.

    \item \textbf{Data Collection:} Data from all nodes are continuously collected during the experiments. This includes positional data, and motion data, providing a comprehensive dataset for analysis.
\end{itemize}

%This experimental setup is integral to developing and refining our proximity-based wearable computing testbed. It enables a thorough evaluation of the system's functionality, reliability, and potential for real-world applications.

\subsection{Access to Generated Performance Data and Measurements}

The data collected by our server plays a crucial role in assessing the testbed's performance. We have implemented a systematic data collection, storage, and analysis approach, focusing on key performance metrics.

\subsubsection{Data Collection and Storage}
Data from the testbed is saved on the server in CSV format, facilitating easy offline analysis. This format allows for a straightforward integration with data analysis tools and libraries.

%\subsubsection{Data Analysis using Python}
%The collected data is analyzed using Python, leveraging the '\textit{pandas}' library for efficient data manipulation and analysis.
%The specific Python code used for analysis varies depending on the performance metric or aspect of the evaluated system.

\subsubsection{Performance Metrics}
\begin{itemize}
    \item \textbf{Accuracy of the Tracking System:} Tests are conducted to evaluate the precision of the localization system. This involves the strategic placement of anchors and collecting data from tags over a certain period.
    
    \item \textbf{Packet Loss Ratio (PLR):} The Packet Loss Ratio of MQTT messages from tags to the server is measured, providing insights into the reliability of the communication system.
    
    \item \textbf{Delay of MQTT Messages:} Analysis of the delay in MQTT messages from tags to the server helps assess the system's responsiveness.
\end{itemize}

\subsubsection{Testing Conditions and Stress-Tests}
Tests are conducted under various operating conditions, such as with all online edge nodes active or only one online edge node, to assess system performance under different scenarios.
    
Stress tests for PLR and delay metrics are performed using specially developed emulation software configured to simulate different load and network conditions.

\subsection{Evaluation of the Testbed}
\subsubsection{Accuracy of the tracking system}
\paragraph{Static Checkpoint Analysis}

\begin{figure} [!ht]
    \centering
    \includegraphics[width=0.9\columnwidth]{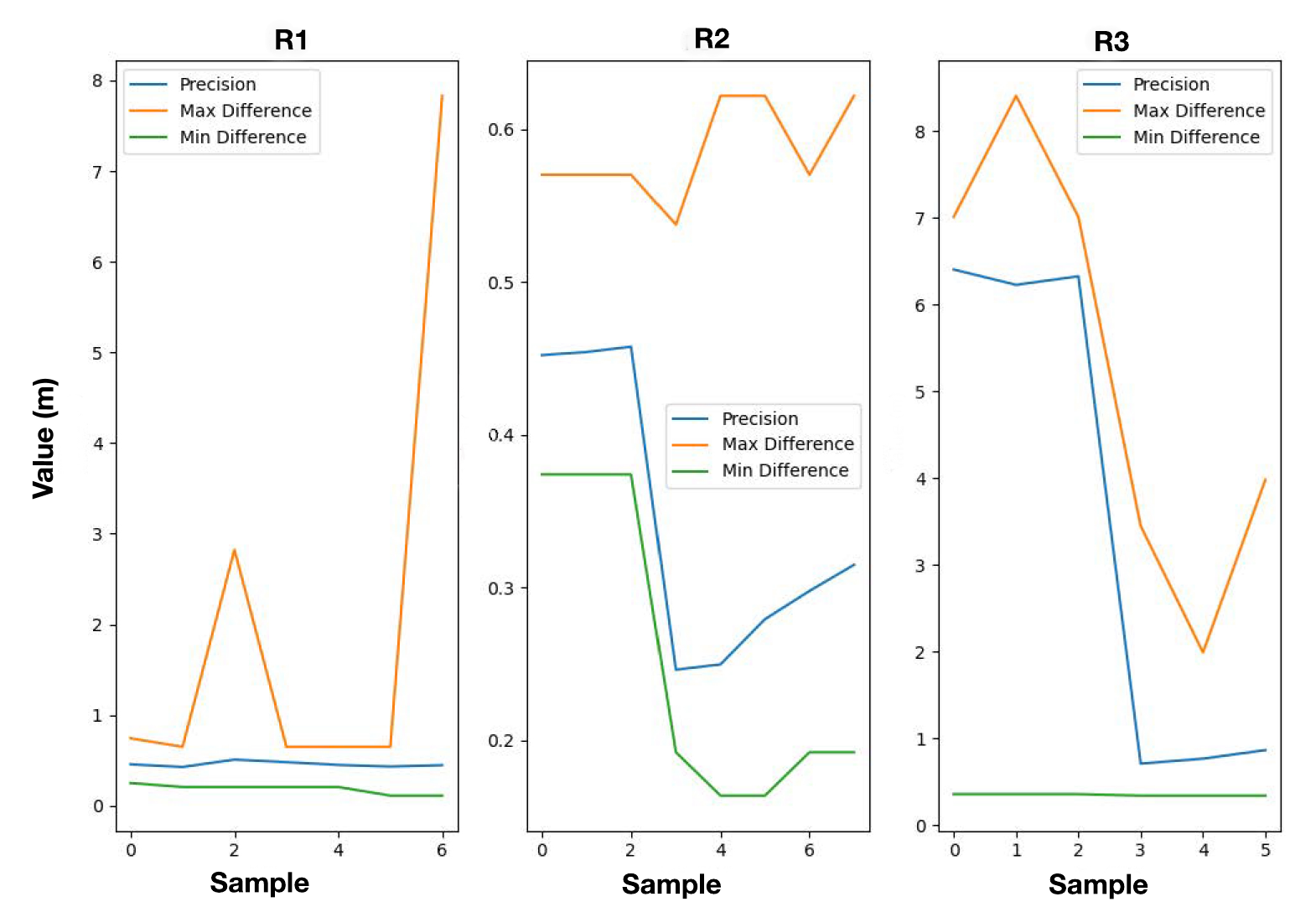}
    \caption{Static Check Point.}
    \label{fig:staticcheckpoint}
\end{figure}

The analysis of static checkpoints, as depicted in Figure 2, showcases the operational performance of the testbed under various conditions. The three checkpoints, R1, R2, and R3, correspond to designated static points P1, P2, and P3. Their coordinates are detailed as follows: P1 at (1.55, 0), P2 at (1.9115, 5.938), and P3 at (5.353, 2.724), represented by red dots in the lower right of the figure.

At checkpoint R1, with recorded data points (1.65, 0), we observe a notable maximum difference combined with lower average precision. This discrepancy may be attributed to intermittent disturbances that compromise data accuracy. However, the minimal difference at R1 suggests a potential for high precision in stable conditions. Checkpoint R2, recorded at (1.9115, 6.038), displays the lowest average precision but a reduced maximum difference, indicating consistent yet less precise readings. The narrow deviation range at R2 signifies stable system behavior, albeit with a consistent margin of error. Finally, R3, located at (5.453, 2.724), demonstrates both higher average precision and a larger maximum difference. This variability could be due to environmental changes or inherent fluctuations within the system, requiring further analysis to understand its impact on performance.

In summary, when outlier data is excluded, the testbed's positioning accuracy is approximately 0.2m. However, this is influenced by the chosen localization algorithm and the spatial arrangement of the sensors, as our setup utilizes two-dimensional coordinate positioning. The results underscore the system's functional performance in static tests, with the precision of localization closely tied to sensor placement and algorithmic efficiency.

\paragraph{Dynamic Checkpoint Analysis}
\begin{figure} [!ht]
    \centering
    \includegraphics[width=0.9\columnwidth]{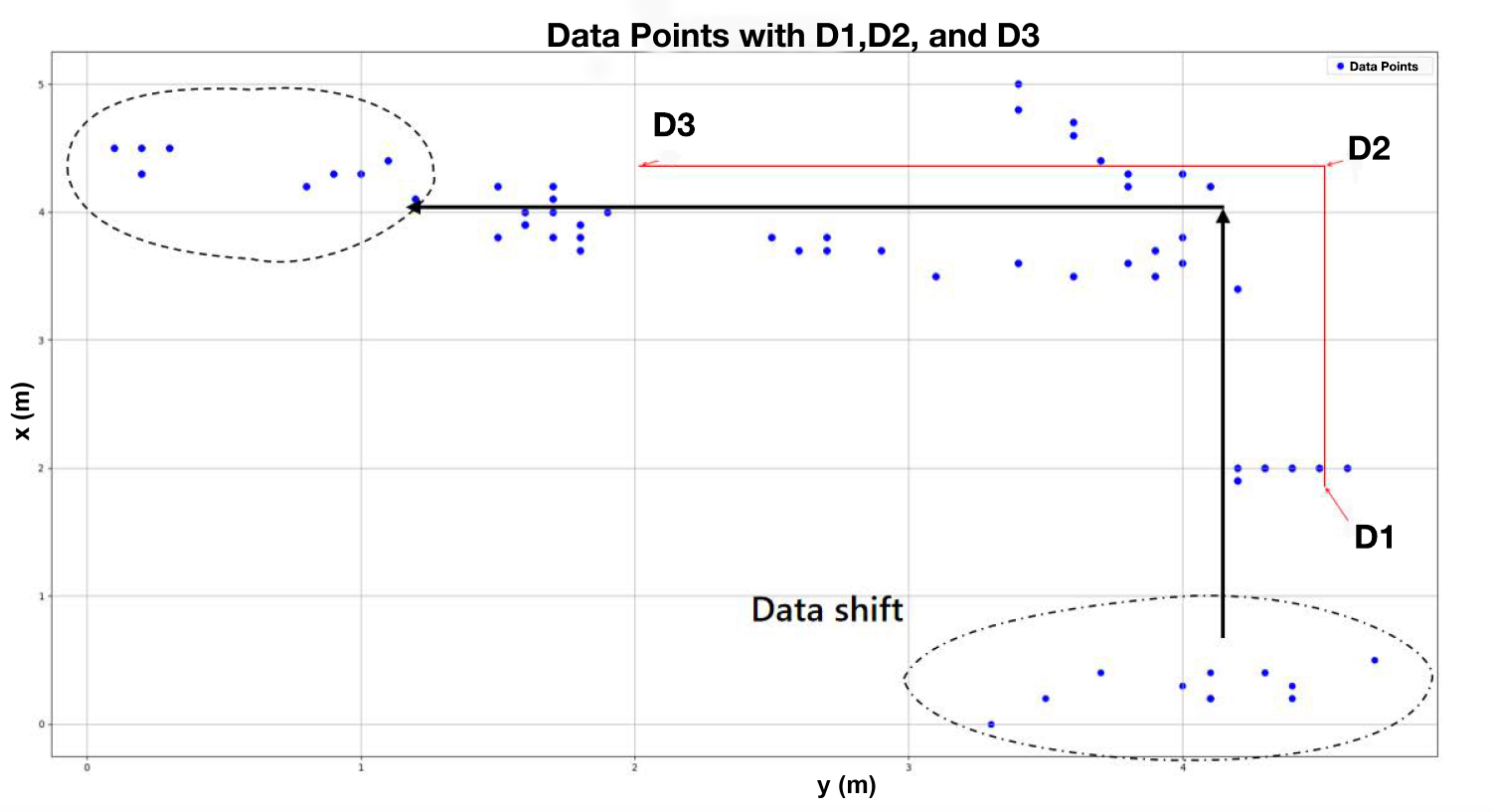}
    \caption{Dynamic Check Point.}
    \label{fig:dynamiccheckpoint}
\end{figure}

In the dynamic testing phase (see Figure~\ref{fig:dynamiccheckpoint}), subjects were directed to move through three fixed checkpoints—D1, D2, and D3—over a span of approximately 5 meters within an 8-second interval. The scatter plot analysis indicates a similar precision level as observed in static conditions, affirming the system's reliability for tracking in motion. Concentrations of data points around the checkpoints suggest well-defined subject pauses or slow movement at these locations. Notably, a lateral data shift observed between D2 and D3 raises questions about potential systematic deviations or environmental influences that were not evident in the static test. This deviation merits further investigation to discern its cause, be it sensor drift, subject movement patterns, or an aspect of the test environment. 

\subsubsection{Testbed Packet Loss Ratio Analysis}

\begin{figure}[ht]
\centering
\includegraphics[width=0.7\columnwidth]{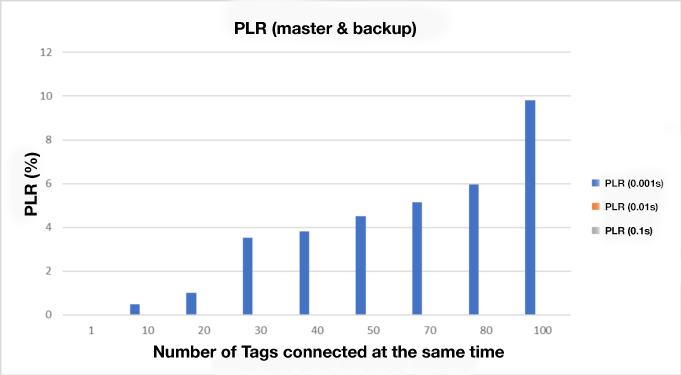}
\caption{PLR with Master \& Backup servers online.}
\label{fig:plr_1}
\end{figure}

\begin{figure}[ht]
\centering
\includegraphics[width=0.7\columnwidth]{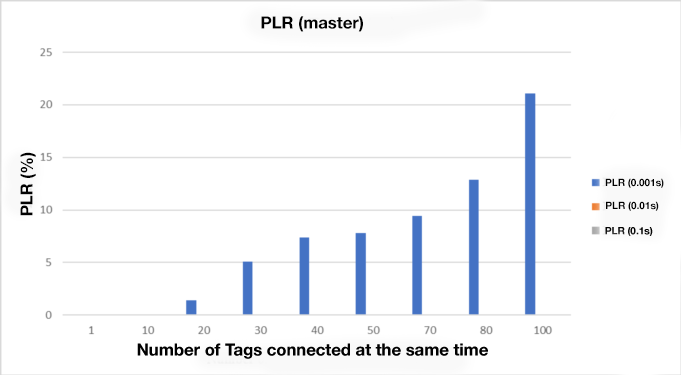}
\caption{PLR with only Master server online.}
\label{fig:plr_2}
\end{figure}

The Packet Loss Ratio (PLR) is a critical performance metric in evaluating the reliability of our testbed's communication system. The PLR was analyzed under two configurations (see Figure~\ref{fig:plr_1},~\ref{fig:plr_2}): with both master and backup servers online and with only the master server online. The data indicates that the dual-server configuration consistently yields a lower PLR across various message frequencies, particularly as the number of simultaneously connected tags increases. This trend demonstrates the effectiveness of a redundant server setup in managing network traffic and reducing packet loss. Conversely, while adequate at lower tag counts, the single-server configuration shows a marked increase in PLR as the frequency of messages and the number of tags increase, culminating in a notable peak when 100 tags are connected at the highest message frequency. These findings highlight the necessity of server redundancy to maintain data integrity, especially in high-load scenarios where real-time data transmission is paramount.

\begin{figure}[ht]
\centering
\includegraphics[width=0.7\columnwidth]{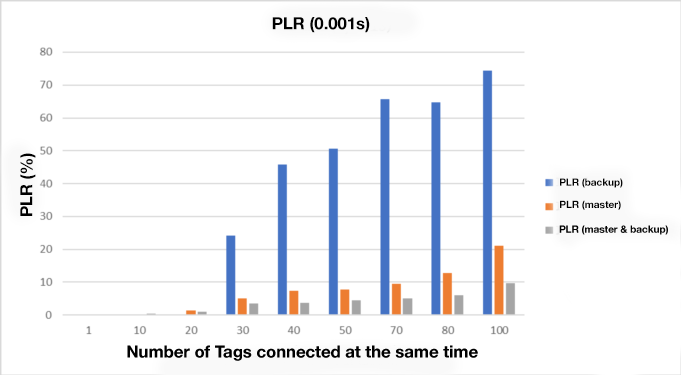}
\caption{PLR dynamic test.}
\label{fig:plr_3}
\end{figure}

The PLR during dynamic testing, as shown in Figure~\ref{fig:plr_3}, particularly at a high message frequency of 1000Hz, presents a comparative view of the network's robustness under different server configurations. The PLR noticeably escalates with increasing tags connected simultaneously, reaching its peak when 100 tags are active. This trend is evident across all server setups, but the dual-server configuration (master and backup) consistently outperforms the single-server setups. It maintains a lower PLR, indicating superior handling of network traffic and resilience to packet loss. The dynamic test results further underscore the importance of redundancy in server architecture to accommodate high-frequency data transmission and a large number of client nodes, as reflected in the significant difference in PLR between the 'master \& backup' versus 'only master' or 'only backup' configurations. 

\subsubsection{Delay Analysis from Tags to Server}
In-depth analysis of communication delay from tags to the server is critical to assessing the testbed's responsiveness and reliability under dynamic conditions. This analysis encompassed a variety of server configurations, including scenarios where only the master server was active, only the backup was operational, and both servers were running concurrently. 

Both IMU and UWB message types, representing different data transmission protocols and bandwidth requirements, were taken into consideration. The tests were conducted across a progressively increasing number of connected tags, starting from a single tag to a dense network of one hundred tags. This range was chosen to simulate varying operational load levels and understand the impact of scale on the system's latency.

The delay measurements were meticulously recorded, focusing on the time it takes for a message to travel from the tag to the server. These times indicate the system's ability to handle real-time data transmission—a critical requirement for applications such as location tracking and environment monitoring.

\begin{figure}[ht]
\centering
\includegraphics[width=0.9\columnwidth]{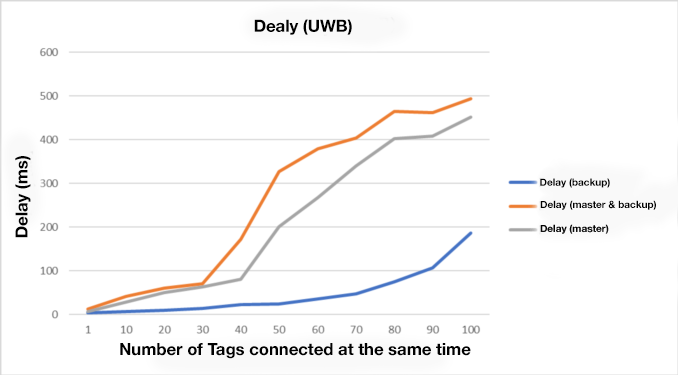}
\caption{Delay (UWB) with varying numbers of tags connected simultaneously.}
\label{fig:delayuwb}
\end{figure}

\begin{figure}[ht]
\centering
\includegraphics[width=0.9\columnwidth]{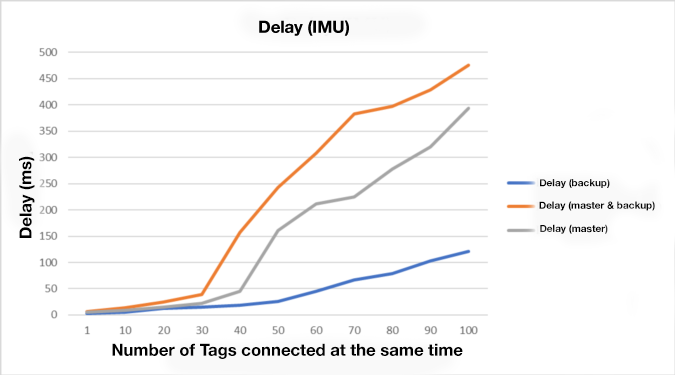}
\caption{Delay (IMU) with varying numbers of tags connected simultaneously.}
\label{fig:delayimu}
\end{figure}

%\subsection{Limitations and Future Directions}
%Our testbed, while innovative in enhancing community interaction, faces challenges like environmental sensitivity of UWB sensors and data security concerns in edge computing. Future efforts will be concentrated on enhancing UWB sensor resilience through advanced algorithms and strengthening data security. Additionally, expanding interoperability with diverse IoT devices and applying the testbed in varied community settings are critical goals. These advancements aim to refine the system's efficacy in wearable computing, offering broader applications in community-centric environments.

\section{Conclusion}

The testbed we have developed stands as a testament to the potential for revolutionizing community interactions, integrating UWB position sensors and 9-axis motion sensors, offering precise proximity detection and intricate movement analysis. While promising, the system grapples with challenges such as environmental susceptibilities that affect sensor reliability and concerns surrounding the security and privacy of edge-computed data. As we look forward, our commitment is to fortify the resilience of our sensors against environmental variables and to bolster our security framework, ensuring data integrity and user privacy are never compromised. Additionally, expanding the testbed's interoperability with a broader range of IoT devices and platforms remains a priority, making it an integral part of a smart, connected community infrastructure. This ongoing journey underscores our dedication to advancing wearable technologies and proximity-based systems for enhanced community integration and interaction.

\section*{Acknowledgment}
This work has been partially supported by the Italian MUR, PRIN 2022 Project “COCOWEARS” (A framework for COntinuum COmputing WEARable Systems), n. 2022T2XNJE, CUP: H53D23003640006 and PRIN 2020 Project “COMMON-WEARS”. % TODO CUP and n. of common-w
This work also received funding by the European Community’s Horizon Europe Programme under the “MLSysOps Project” (Grant Agreement 101092912).

\bibliographystyle{IEEEtran}
\bibliography{references.bib}

\flushend
\end{document}